\documentclass[final,12pt,a4paper]{elsart}
\usepackage{graphics}
\usepackage{epsfig}
\usepackage{subfigure}
\usepackage{amssymb}
\usepackage{amsmath}

\begin{document}
\begin{frontmatter}
\title{Microscopic theory of the activated behavior of the quantized Hall effect}

\author[l1]{S. Sakiroglu},
\author[l2]{U. Erkarslan},
\author[l2]{G. Oylumluoglu},
\author[l2]{A. Siddiki}
and
\author[l1]{I. Sokmen}
\address[l1] {Physics Department, Faculty of Arts and Sciences, Dokuz
Eyl{\"u}l University\\ 35160 \.{I}zmir, Turkey}
\address[l2]{Physics Department, Faculty of Arts and Sciences,
Mu{\v{g}}la University\\ 48170 Mu{\v{g}}la, Turkey}

\begin{abstract}
The thermally activated behavior of the gate defined narrow Hall
bars is studied by analyzing the existence of the incompressible
strips within a Hartree-type approximation. We perform
self-consistent calculations considering the linear response
regime, supported by a local conductivity model. We investigate
the variation of the activation energy depending on the width of
samples in the range of $2d\sim [1-10]~\mu m$. We show that the
largest activation energy of high-mobility narrow samples, is at
the low field edge of Hall filling factor 2 plateau (exceeding
half of the cyclotron energy), whereas for relatively wide samples
the higher activation energy is obtained at the high field edge of
Hall plateau. In contrast to the single-particle theories based on
the localization of electronic states, we found that the
activation energy is almost independent of the properties of the
density of states.
\end{abstract}


\end{frontmatter}
A two-dimensional electron system (2DES) subjected to strong
magnetic fields perpendicular to the 2DES exhibits the integer
quantum Hall effect (IQHE)~\cite{Klitzing80}. Due to the extremely
high reproducibility of the certain quantized resistance values,
QHE stands as a resistance standard~\cite{Cage84}. The appearance
of narrow resistivity peaks separated by deep minima is a defining
feature of the QHE \cite{Klitzing05}. The standard explanation of
the IQHE is based on single-particle picture which accepts the
Landau quantization \emph{and} localization of electronic states
as key points \cite{Svoboda97}. The Coulomb interaction between
the electrons is neglected in this picture and is unable to
explain several important features, such as high reproducibility
and precise quantization \cite{Siddiki04,Siddiki08_1}. Recently, a
microscopic interpretation of the IQHE that incorporates the
electron-electron interaction explicitly \cite{Gerhardts_arxiv06},
provides a prescription to calculate the Hall and longitudinal
resistances ($R_{xy}$ and $R_{xx}$, respectively) explicitly under
experimental conditions. It is stated that a quantitative theory
that describes the activated behavior has to include an analysis
of the longitudinal and Hall conductivity and their dependencies
on the temperature, magnetic field, current density, sample widths
and the other material properties \cite{Matthews05}. There have
been several attempts to explain the physics behind the key
features of activation of the quantum Hall effect. For different
temperature regimes various transport processes have been proposed
\cite{Rubinger06,Svoboda_arxiv}. At intermediate temperatures
($10$ K$<T<20$ K) conductance is predominantly determined by
thermal activation of electrons. The temperature dependence of the
conductivity $\sigma_{xx}$ is thermally activated with an
activation energy $E_{a}$ corresponding to the energy difference
between the Fermi energy and the mobility edge \cite{Furlan98}. In
the single particle theories, it is expected that the largest
activation energy is obtained if the Fermi energy resides at the
midpoint between two Landau levels \cite{Svoboda97}, \emph{i.e.}
at the center of the $B$ field interval where $R_{xx}$ vanishes.
However, in the literature strong deviations are reported when
considering high mobility narrow samples
~\cite{Siddiki08_1,Tsui85}

The purpose of the present paper is to summarize the results on
activated behavior of high-mobility Hall bars studied within a
microscopic theory which avoids any localization assumptions. A
gate defined narrow Hall bar system with sample width $2d$ is
constructed by in-plane metallic side gates kept in zero
potential, \emph{i.e.} $V(-d)=V(d)=0$ dictating the boundary
conditions to the solution of the relevant Poisson equation. The
2DES is depleted from the edges by an amount of $|b|/d$ which
resides in the $z=0$ plane. The spatial distribution of the
electron number density in $x-$ direction is denoted by $n_{\rm
el}(x)$ and translational invariance in $y-$direction is assumed.
Electron system is populated by ionized Si-donors with average
density $n_{0}$ which are distributed homogeneously in the same
plane with the 2DES. The confinement potential $V_{\rm bg}(x)$
determined by these background charges is obtained from the
solution of the Poisson equation yielding the kernel
\begin{equation}
K(x,x^{\prime})=\ln{\left|\frac{\sqrt{(d^{2}-x^{2})(d^{2}-{x^{\prime}}^{2})}+d^{2}-x\,x^{\prime}}{(x-x^{\prime})d}\right|},
\end{equation}
via
\begin{equation}
V_{\rm
bg}(x)=\frac{2e^2}{\bar{\kappa}}\int\limits_{-d}^{d}dx^{\prime}K(x,x^{\prime})n_{0},
\end{equation}
which leads to
\begin{equation}
V_{\rm bg}(x)=-E_{0}\sqrt{1-(x/d)^{2}}, \qquad
E_{0}=2\pi\,e^{2}n_{0}d/\overline{\kappa}.
\end{equation}
In above equations $\overline{\kappa}$ is an average background dielectric
constant of the material. The Hartree potential due to the Coulomb
interaction between the electrons is obtained by using the same
electrostatic kernel as
\begin{equation}
V_{\scriptsize{\rm
H}}(x)=\frac{2\,e^{2}}{\overline{\kappa}}\int\limits_{-d}^{d}dx^{\prime}K(x,x^{\prime})n_{\rm
el}(x^{\prime}).
\end{equation}
Hence the electrons are subjected to an effective potential
defined as
\begin{equation}
V(x)=V_{\rm bg}(x)+V_{\scriptsize{\rm H}}(x)\label{eq:potential}.
\end{equation}
We neglect the antisymmetry condition for the Fermionic wave
functions in calculation of electron density, considering spinless
particles. Such simplification is justified due to the small
$g^*$-factor ($\approx -0.44$) of a 2DES induced on a GaAs/AlGaAs,
the Zeeman energy is much smaller than the Landau energy. Since
the confining potential for electrons varies smoothly over the
extend of the eigenfunction. We employ the self-consistent
Thomas-Fermi Approximation which describes realistically the
electronic distribution as,
\begin{equation}
n_{\rm el}(x)=\int dE D(E)f(E),\label{eq:density}
\end{equation}
where $D(E)$ is the density of states described by self-consistent
Born approximation~\cite{Ando_82} in the presence of strong
magnetic field and $f(E)$ is the Fermi-Dirac distribution function
which is position dependent via the local electrostatic
potential~\cite{Siddiki04}. After solving eq.~\ref{eq:potential}
and eq.~\ref{eq:density} self-consistently, the current distribution
is obtained by utilizing the local version of the conventional
transport theory, i.e. the Ohm's law, which takes into account
implicitly the peculiar screening effects in 2DES under the high
magnetic fields, however avoids any localization assumptions. The
nonlocal effects on the conductivities, adopted from the results
of self-consistent Born approximation, are simulated by
coarse-graining the conductivity tensor according to
\begin{equation}
\hat{\overline{\sigma}}=\frac{1}{2\lambda}\int\limits_{-\lambda}^{\lambda}d\chi\hat{\sigma}(x+\chi),
\end{equation}
where $\lambda=\lambda_{\rm F}/2$ and $\lambda_{\rm F}$ is the
mean particle distance, \emph{i.e.} Fermi wavelength. Since
without any disorder it is not possible to define the
conductivity, we considered impurity potentials described by a
Gaussian
\begin{equation}
\upsilon({\bf{r}})=\frac{V_{0}}{\pi R^{2}}\exp{\left(-\frac{r^{2}}{R^{2}}\right)}
\end{equation}
with the single particle range $R$ of the order of the spacing
between 2DES and doping layer, where $V_0$ is the impurity
strength with relevant dimensions. A configuration average of such
impurity potentials lead to the spectral function $A_{N}(E)$
defined as
\begin{equation}
A_{N}(E)=\frac{2}{\pi\Gamma_{N}}\sqrt{1-\frac{E-E_{N}}{\Gamma_{N}}}.
\end{equation}
The temperature dependencies of the longitudinal and the Hall
resistances are obtained from the position dependent
two-dimensional resistivity tensor \cite{Ando_82} and are studied
at magnetic field intervals within the integer Hall plateau with
filling factor $\nu=2$ considering high mobility samples.

In this work we consider the extremely clean GaAs/(AlGa)As
heterostructures where local DOS is almost a delta function and no
long-range potential fluctuations exist. We fixed the background
charges to the density $n_{0}=4\times10^{11}\, $cm$^{-2}$ and
obtained the corresponding density profile at zero temperature and
zero field potential. Subsequent step is finding iteratively the
positions of incompressible strips (ISs) for the finite
temperature and finite field, where we fixed the depletion length
to $|b|/d=0.9$. Throughout the work we used the Fermi energy
$E_{F}^{0}$ corresponding to the electron density at the center of
the sample ($n_{\rm el}(0)$) as a reference energy.

\begin{figure}[h!]
\begin{center}
\subfigure[]{\epsfig{file=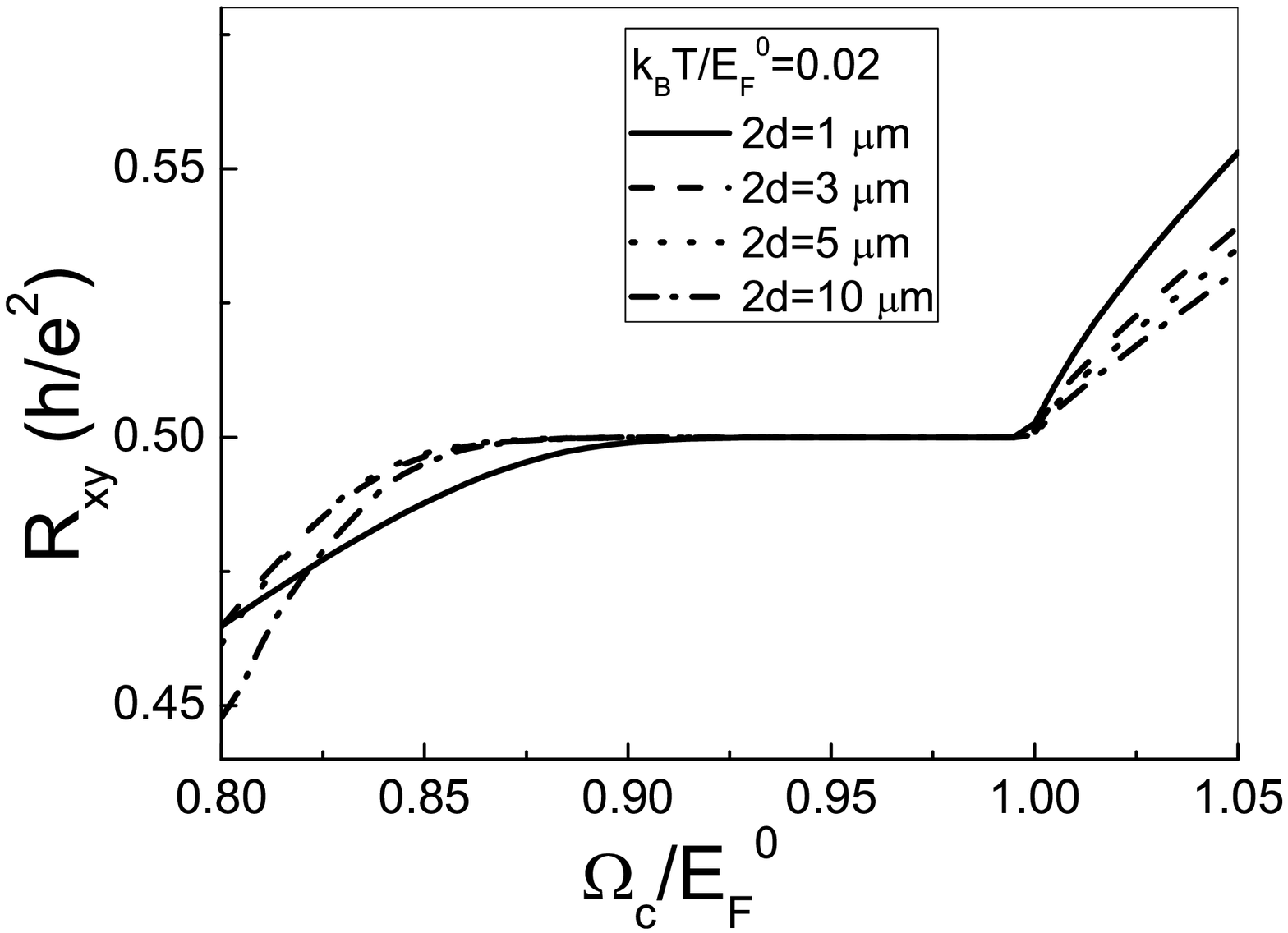,width=6.7cm,height=6.3cm}}
\subfigure[]{\epsfig{file=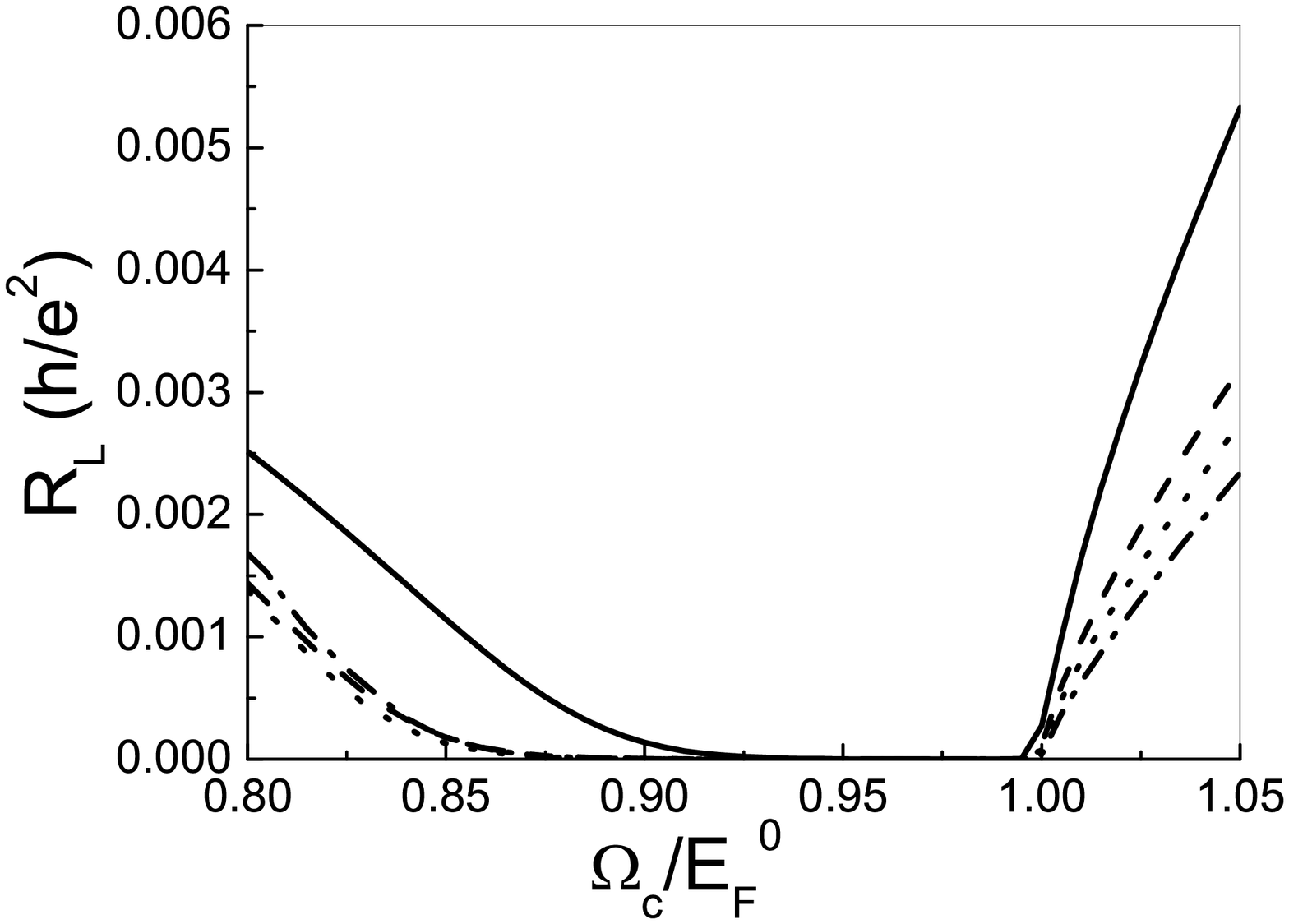,width=6.7cm,height=6.3cm}}
\caption{\label{Fig1:RxyRxx} Calculated Hall (a) and (b)
longitudinal resistances versus scaled magnetic field
$\hbar\omega_{c}/E_{F}^{0}$, with $\omega_c=eB/m^*c$ for different
values of sample widths. The parameters are $R= 20 nm$,
$\Gamma/\hbar\omega_{c}=\Gamma/\Omega_{c}=0.05$, and
$k_{B}T/E_{F}^{0}=0.02$.}
\end{center}
\end{figure}

The numerical results in Fig.~\ref{Fig1:RxyRxx} depicts $R_{xy}$
and $R_{xx}$ as a function of magnetic field. Whenever the system
is in the plateau regime we conclude that an IS of filling factor
$\nu=2$ is formed \emph{somewhere} across the Hall bar, since
backscattering is absent.

\begin{figure}[!h]
\begin{center}
\subfigure[$2d= 1\mu
m$]{\epsfig{file=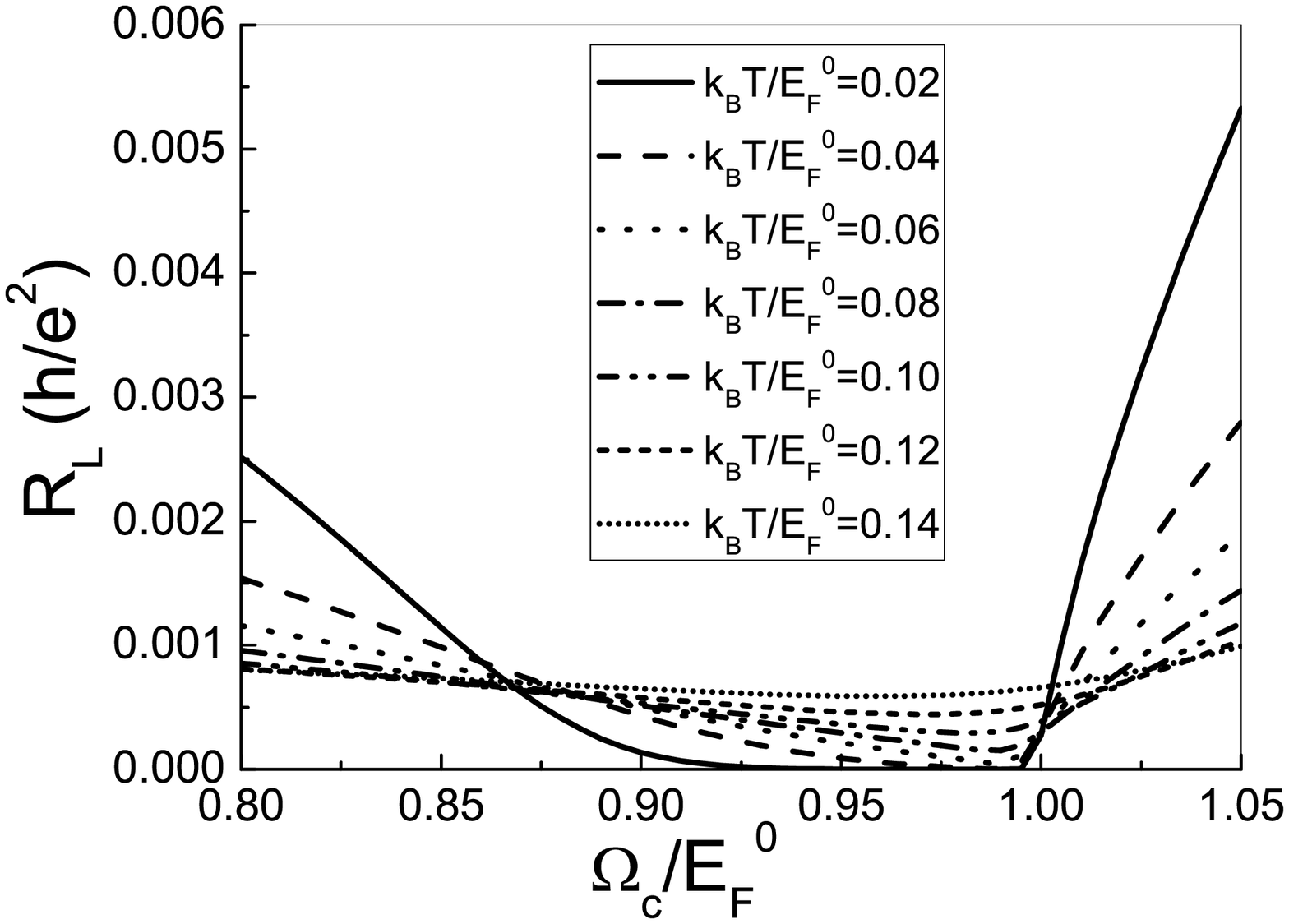,width=6.7cm,height=6.3cm}}
\subfigure[$2d= 10\mu
m$]{\epsfig{file=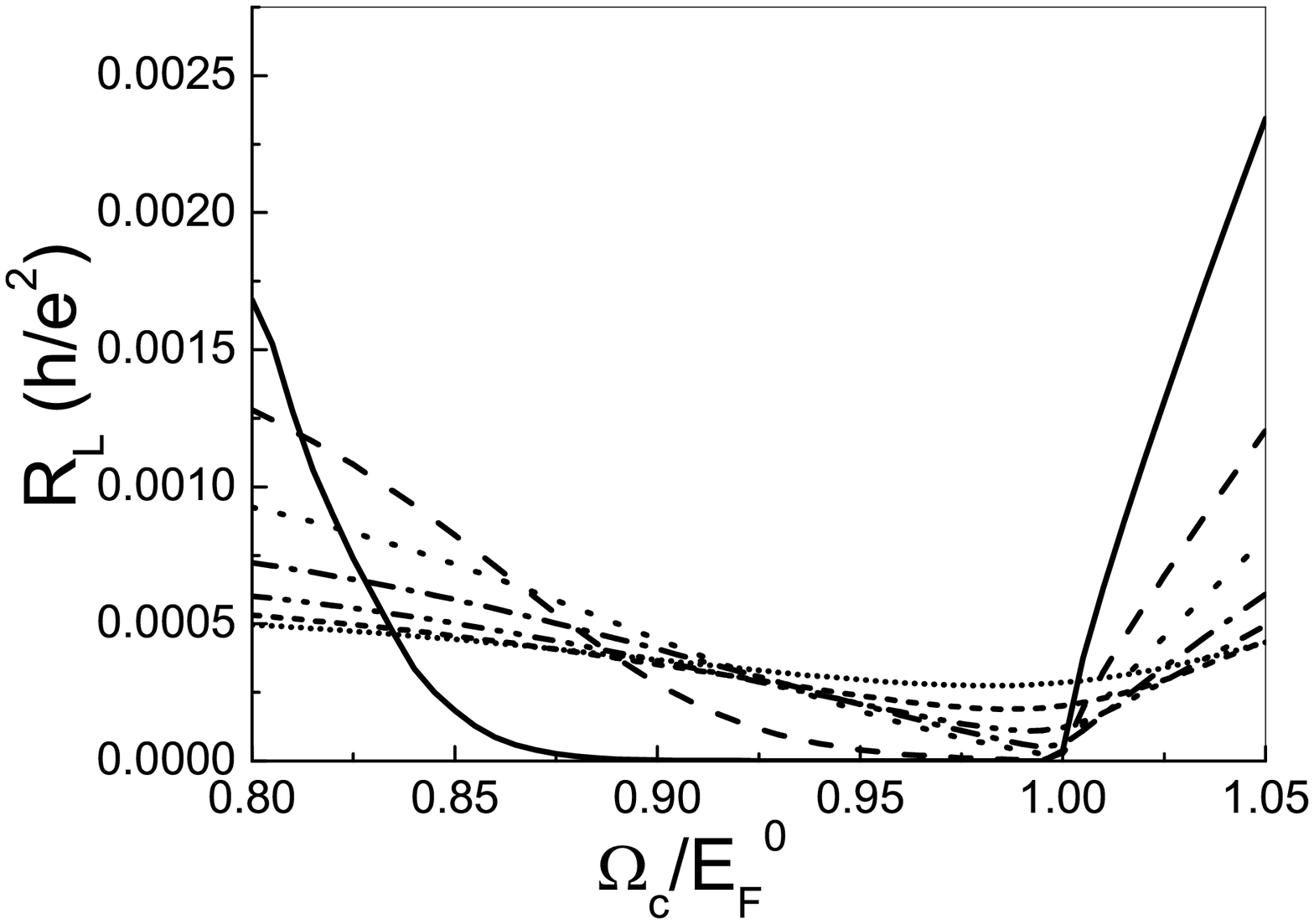,width=6.7cm,height=6.3cm}}
\caption{\label{Fig2:RxxkT} Dependency of the calculated $R_{xx}$
as a function of magnetic field for different temperatures.}
\end{center}
\end{figure}

Within this strip the electrostatic potential varies by the amount of a cyclotron energy and current
flows only in this IS, whereas the adjacent regions are
compressible strips where nearly perfect screening occurs and
partially filled Landau level is pinned to the Fermi energy. Once
the ISs become leaky, \emph{i.e.} if the strip widths become
narrower than the averaging length, scattering between $\nu>2$ and
$\nu<2$ compressible states is possible, then the quantized Hall
effect disappears. Standard Hall wafers contain a 2DES of typical density $n_{\rm
el}(0)\leq 4\times 10^{11}\,$cm$^{-2}$, hence we find a finite
magnetic field interval of finite in which IS with integer value
of the local filling factor 2 exist. For the B values in this
interval, the deviation of the Hall resistance from the quantized
values increases with increasing temperature. To calculate the
activation energy it is crucial to identify the exact B value for
$\bar{\nu}=2$, where bar stands for the average filling factor.
For standard Hall bars, where edge effects are suppressed by the
disorder effects and the activation energy is calculated at the
center of the plateau, i.e $\bar{\nu}=2$. In contrast, for narrow
samples where edge effects are predominant and electron
distribution is no longer flat. Since the behavior of the crossing
of classical curve and the low temperature curve is asymmetric
with respect to the center of the plateau, it is not
straightforward to locate the B value corresponding to
$\bar{\nu}=2$. In order to define the B value where the activation
energy should be calculated, we present the temperature dependence
of the longitudinal resistance for narrow and wide samples in
Fig.~\ref{Fig2:RxxkT}.

The relevant B value is determined where the longitudinal
resistance remains minimal while increasing the temperature. For
the $2d=1~\mu $m this critical $B$ value is found to be at
$B_a=\Omega_{c}/E_{F}^{0}=0.990$ while for $2d=10~\mu$m at
$B_a=\Omega_{c}/E_{F}^{0}=0.995$.

\begin{figure}[!h]
\begin{center}
\subfigure[]{\epsfig{file=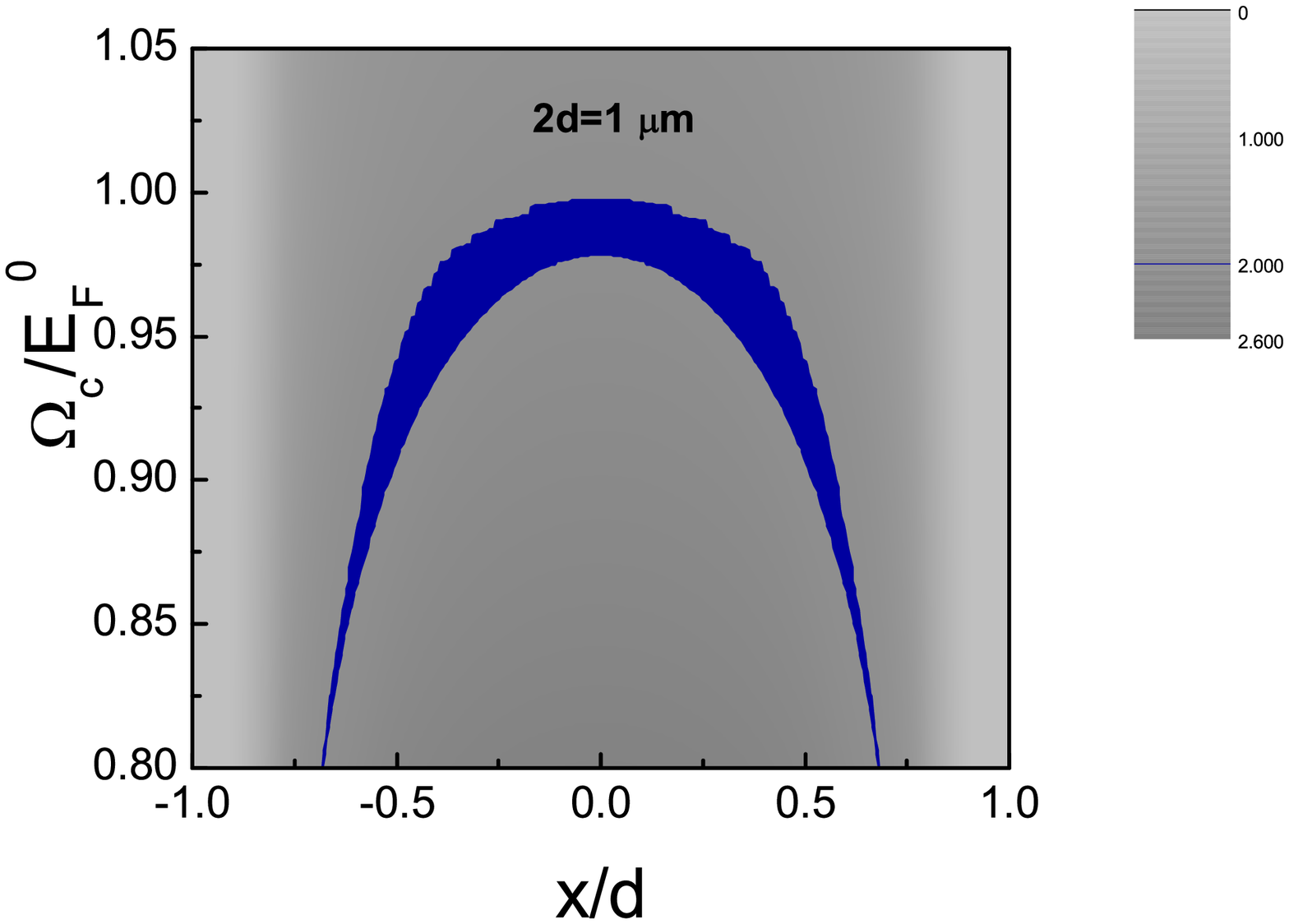,width=6.7cm,height=6.0cm}}
\subfigure[]{\epsfig{file=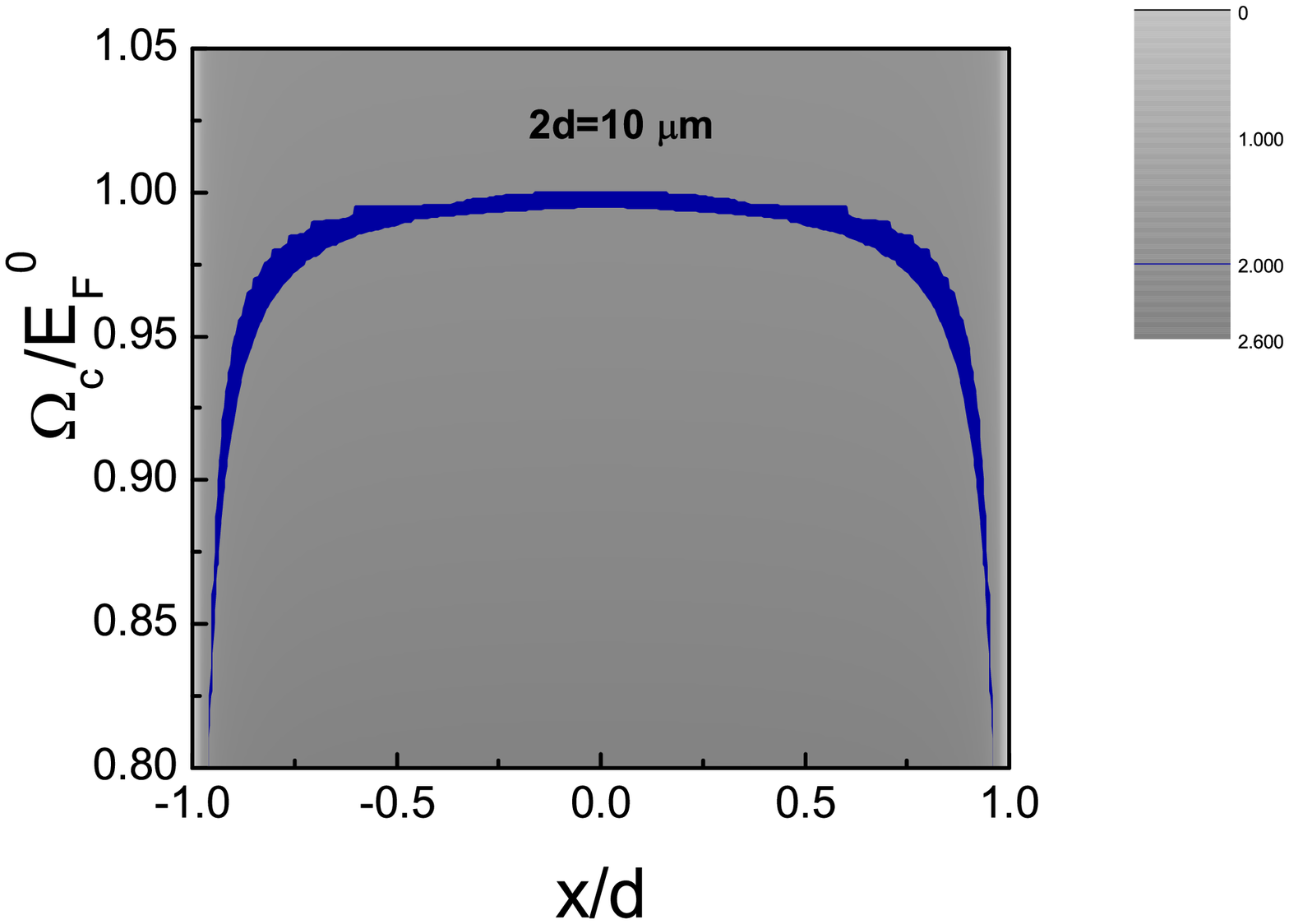,width=6.7cm,height=6.0cm}}
\caption{\label{Fig3:Hilal} Gray scale plot of the averaged
filling factor profile versus position $x$ and magnetic field
$\Omega_{c}/E_{F}^{0}$ at $k_{B}T/E_{F}^{0}=0.02$. The regions of
IS with $\nu=2$ are indicated.}
\end{center}
\end{figure}

In order to clarify the relation between the formation of the ISs
and the global resistances, we present the calculated local
filling factor profile as a function of the scaled lateral
coordinate and varying magnetic field in Fig.~\ref{Fig3:Hilal}, as
gray scale. At sufficiently large $B$ field $\Omega_{c}/E_{F}^{0}>1$, the
local filling factor $\nu(x)$ is less than 2 everywhere and system
is completely compressible.

Reducing the $B$ leads to the
formation of IS at the center of the sample with $\nu(x)=2$ while
gradually decreases to zero toward the edge of the sample. Further
decrease of $B$, enforces the ISs with $\nu=2$ move toward the
sample edge with narrowing its width, meanwhile filling factor of
the central region increases.

\begin{figure}[!t]
\begin{center}
\epsfig{file=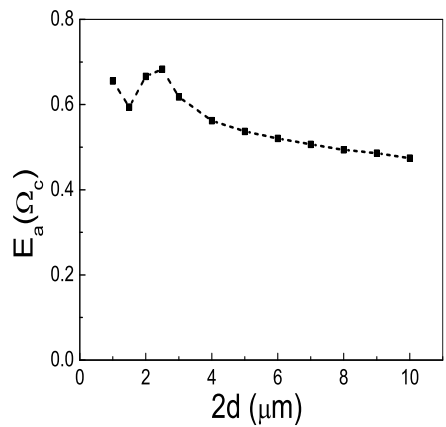,width=7.0cm,height=6.3cm}
\caption{\label{Fig4:Ead} Dependence of the activation energy of
the $\nu=2$ Hall plateau on the width of the samples.}
\end{center}
\end{figure}

As a next, at low bias currents, thermally activated resistance is
investigated by the Arrhenius plot at the relevant magnetic field
values. The activation energies were extracted from fitting
\begin{equation}R_{xx}(T,B_a)=R_{xx}^{0}\exp{(-E_{a}/2\,k_{B}T)}\nonumber \end{equation}
to the maximum slopes of the data points. Results for $E_{a}$ are
shown in Fig.~\ref{Fig4:Ead}. For relatively wide samples, activation energy is calculated at
the high field edge of Hall plateau whereas for the narrower
samples activation energy is obtained at the low field edge.
Energies for narrower samples exceeds half of the cyclotron energy
of the interval between Landau levels. However with increase in
width of sample the asymptotic decrease in activation energy
recovers the well known values. The non monotonic behavior
observed at narrower samples $2d<2~\mu$m is closely related with
the formation of the large bulk incompressible region. Such
anomalies are discussed elsewhere~\cite{bizimPRB}.

In summary, the temperature dependence of the longitudinal
conductivity has been studied in the high-mobility gate defined
narrow Hall bar samples of with well developed IQHE plateaux in
magnetic field interval corresponding to filling factor $\nu=2$.
Activation energies obtained by fitting the data to the Arrhenius
law are calculated at the magnetic field where the longitudinal
resistance remains the smallest with increase in temperature.

In contrast to the single-particle theories, we found that the
activation energy depends strongly on the width of sample. The
highest values of activation for narrow samples are obtained at
the low field edge of Hall plateau whereas for wider samples this
values shifts to the high field edge. Activation energies, for
extremely narrow samples exceeding the half of the cyclotron
energy, decreases asymptotically with increase of sample width. We
suggest that the enhanced contribution to the activation energy is
connected with the width of ISs which promotes the thermally
activated conduction.

\end{document}